\documentclass[twocolumn,showpacs,aps,prl,floatfix,superscriptaddress,nofootinbib]{revtex4-2}
\usepackage{graphicx}
\usepackage{epstopdf}
\usepackage{color}
\usepackage{amsmath}
\usepackage{hyperref}
\usepackage{url}

\begin{document}

\preprint{APS/123-QED}

\title{Controlling Spin-Waves by Inhomogeneous Spin-Transfer Torques}
\author{Lorenzo Gnoatto}
\email{l.g.gnoatto@tue.nl}
\affiliation{Department of Applied Physics and Science Education, Eindhoven University of Technology, P.O. BOX 5132, 5600 MB Eindhoven, The Netherlands\\}
    
\author{Jean F. O. da Silva}
\affiliation{COMMIT, Department of Physics, University of Antwerp, Groenenborgerlaan 171, B-2020 Antwerp, Belgium\\}

\author{Artim L. Bassant}
\affiliation{Institute for Theoretical Physics, Utrecht University, Princetonplein 5, 3584 CC Utrecht, The Netherlands\\}

\author{Rai M. Menezes}%
\affiliation{COMMIT, Department of Physics, University of Antwerp, Groenenborgerlaan 171, B-2020 Antwerp, Belgium\\}
\affiliation{Departamento de Física, Universidade Federal de Pernambuco, Cidade Universitária, 50670-901 Recife-PE, Brazil\\}

\author{Rembert A. Duine}
\affiliation{Department of Applied Physics and Science Education, Eindhoven University of Technology, P.O. BOX 5132, 5600 MB Eindhoven, The Netherlands\\}
\affiliation{Institute for Theoretical Physics, Utrecht University, Princetonplein 5, 3584 CC Utrecht, The Netherlands\\}

\author{Milorad V. Milo\v{s}evi\'c}%
\email{milorad.milosevic@uantwerpen.be}
\affiliation{COMMIT, Department of Physics, University of Antwerp, Groenenborgerlaan 171, B-2020 Antwerp, Belgium\\}

\author{Reinoud Lavrijsen}
\affiliation{Department of Applied Physics and Science Education, Eindhoven University of Technology, P.O. BOX 5132, 5600 MB Eindhoven, The Netherlands\\}

\begin{abstract}
We investigate the interplay between spin currents and spin waves in nanofabricated Permalloy waveguides with geometrical constrictions. Using propagating spin-wave spectroscopy, micromagnetic simulations, and analytical modeling, we provide experimental evidence that spin-wave phase can be modulated by inhomogeneous spin-transfer torques generated by current-density gradients shaped by the constriction geometry. Narrower constrictions enhance these gradients and modify the internal field for Damon–Eshbach spin waves, resulting in pronounced changes in spin-wave group velocity and phase. To our knowledge, this constitutes the first demonstration of deterministic phase modulation via engineered nonuniform spin-transfer torques. Beyond enabling a scalable route to magnonic interferometry—a building block for spin-wave-based computing—our findings establish a platform to control spin-wave dynamics in spatially varying current landscapes, relevant for analogue-gravity experiments in condensed matter systems.
\end{abstract}



\date{\today}

\maketitle



The interaction between spin-polarized currents and magnons has emerged as a powerful mechanism for manipulating spin waves (SWs) in magnonic systems. This coupling, mediated by spin-transfer torque (STT)~\cite{SLONCZEWSKI1996L1}, enables control over SW amplitude~\cite{gladii2016spin}, frequency~\cite{SW_doppler_exp}, and, as in this work, phase. Early theoretical and experimental studies~\cite{SLONCZEWSKI1996L1,PhysRevB.54.9353,Nist_review} focused on uniform direct currents, revealing effects such as the spin-wave Doppler shift~\cite{SW_doppler_exp} and establishing the foundations of STT physics~\cite{PhysRevB.70.024417,PhysRevLett.93.127204}. In this context, SW magnetization dynamics is subjected to a STT-induced frequency shift given by \(\Delta\omega_{\mathrm{STT}} = -v_{\mathrm{d}} k\)~\cite{SW_doppler_exp}, where the spin-polarized electron drift velocity is \(v_{\mathrm{d}} = gP\mu_{\mathrm{B}}J / (2|e|M_{\mathrm{s}})\) and $k$ is the wavevector. Here, \(g\) is the Landé \(g\)-factor, \(P\) the spin polarization, \(\mu_{\mathrm{B}}\) the Bohr magneton, \(e\) the elementary charge, \(M_{\mathrm{s}}\) the saturation magnetization, \(k\) the wavevector, and \(J\) the current density—explicitly showing the direct proportionality between STT effects and \(J\). Note that the use of spin-orbit-torques (SOTs) \cite{HARMS2022,GLADII-SOT} has also been explored but in first order does not lead to ‘doppler shifts’ as is the aim of our work. 
 
SWs are dispersive waves whose properties are defined by their dispersion relation \(\omega(k)\), which links frequency, wavevector, and other magnetic parameters such as saturation magnetization, external applied magnetic field and effective field~\cite{10.1063/5.0019328, pirro2021advances}. Their group velocity, \(v_{\mathrm{g}} = \partial \omega / \partial k\), sets the speed at which energy and information are transported by the wavepacket, while the phase velocity, \(v_{\mathrm{p}} = \omega / k\), corresponds to the speed of an individual spectral component. Since STT modifies the dispersion, it can affect both \(v_{\mathrm{g}}\) and \(v_{\mathrm{p}}\), with \(v_g\) maximized when the electron flow and wavevector are co-directed. This dual control forms the basis for the STT-driven phase-modulation effects investigated experimentally and theoretically in this work.

Phase modulation enabled by spatially varying spin-polarized currents, could serve as a building block for interferometric devices such as Mach–Zehnder-type spin-wave interferometers~\cite{Mach-Zehnder_SW,Mach-Zehnder_SW_logic}. However, practical implementations remain limited, and the influence of spatially varying spin-polarized currents on propagating SWs is not fully understood. It has been previously shown theoretically that inhomogeneous currents can give rise to exotic dynamical regimes, including magnonic analogues of black holes~\cite{PhysRevLett.118.061301} and spin-wave lasing~\cite{PhysRevLett.122.037203,PhysRevApplied.22.064086}, in regimes where the spin-drift velocity ($v_{d}$) exceeds the spin-wave group velocity ($v_{g}$). While those works focused on extreme transport conditions, the present study addresses a related but less explored question: how current density gradients, even in the linear regime, influence spin-wave propagation and phase. Understanding this regime is essential for realizing analogue gravity scenarios in magnetic systems, where smooth variations in $v_{d}$ play the role analogous at spacetime curvature~\cite{PhysRevLett.118.061301}.

Experimental studies on inhomogeneous current transport have largely focused on spin-Hall nano-oscillators~\cite{demidov2014nanoconstriction}, where current gradients modulate local magnetization dynamics~\cite{demidov2017magnetization}. Yet, their impact on the propagation of spin waves in extended metallic ferromagnets has, to our knowledge, not been explored. In this work, we demonstrate that geometric constrictions, i.e. the cross-section of a Permalloy waveguide strip varies with position leading to a position dependent current density, which in its turn produces a gradients in the STT-induced spin-wave phase modulation via the inhomogeneous STT. Using propagating spin-wave spectroscopy (PSWS)~\cite{lucassen2019optimizing,vlaminck2010spin}, micromagnetic simulations, and analytical modeling, we show that phase accumulation is highly sensitive to constriction geometry, providing a pathway toward controlling spin-wave interference in reconfigurable magnonic circuits.

\begin{figure}[t!]
\centering
\includegraphics[width=\linewidth]{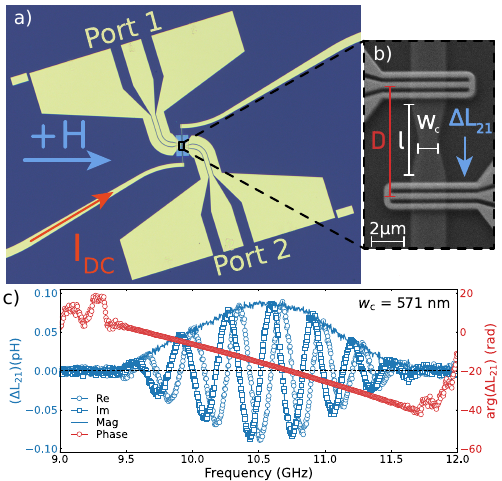}
\caption{(a) Optical microscope image of the PSWS device, showing the orientations of the applied DC current (\( I_{\mathrm{DC}} \)) and the external magnetic field (\( H \)). (b) Scanning electron microscope image of the antenna region and ferromagnetic strip, featuring a constriction with width \( w_{\mathrm{c}} = 1140~\mathrm{nm} \), constriction total tapered length \( l = 4.5~\mu\mathrm{m} \) and antenna distance center to center D=6.1 $\mu$m . (c) Example mutual inductance spectra for a device with \( w_{\mathrm{c}} = 571~\mathrm{nm} \), showing the real, imaginary, and magnitude components of the signal, as well as the corresponding phase (right axis), measured at \( \mu_0 H = +100~\mathrm{mT} \).
}
\label{fig1}
\end{figure}

To probe the impact of inhomogeneous STT on spin-wave transport, PSWS offers access to the frequency~\cite{SW_doppler_exp}, amplitude~\cite{gladii2016spin}, and phase response~\cite{WEISS2023170002} of spin waves with high precision between two points along a spin-wave conduit. Being inherently a two-port measurement~\cite{vlaminck2010spin}, PSWS probes the cumulative effect of spin-wave propagation between the excitation and detection antennas. The devices consist of 2~$\mu$m-wide Permalloy microstrips, fabricated as described in Ref.~\cite{gnoatto2025investigating}, with a Ta(4)/Py(20)/Ta(4) layer stack (thicknesses in nanometers). Fig.~\ref{fig1}(a) shows a wide-field optical microscopy image of a typical PSWS device. To introduce a spatially varying current density, we patterned a central constriction of width $w_\mathrm{c}$ between the two antennas, using $w_\mathrm{c} = 2000$, 1427, 1140, 857, 571, and 285~nm. These values correspond to the nominal widths defined in the lithography design and were arbitrarily chosen. The length ($l$) of the tapered section was kept constant at 4.5 $\mu$m for all devices. Samples with $w_\mathrm{c} < 2000$~nm are referred to as constricted samples. A scanning electron microscopy image of the central region is shown in Fig.~\ref{fig1}(b), indicating the width variation and the directions of spin-wave propagation and applied DC current.

\begin{figure}[t!]
\centering
\includegraphics[width=\linewidth]{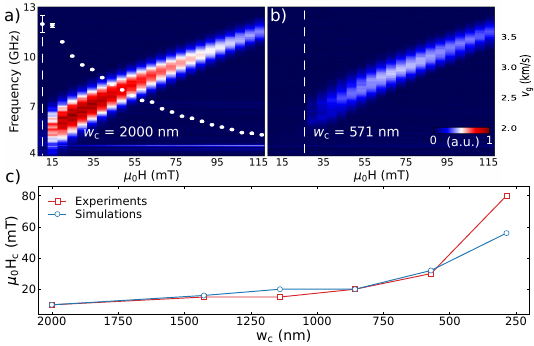}
\caption{Transmitted spin-wave spectra (mutual inductance amplitude) for (a) a uniform waveguide ($w_{\mathrm{c}} = 2000$~nm) and (b) a constricted waveguide ($w_{\mathrm{c}} = 571$~nm). The dots in (a) represent the extracted group velocity, plotted using the right-hand axis shown in panel (b) for clarity.
 (c) Minimum applied magnetic field required for detectable spin-wave transmission, plotted as a function of constriction width. }
\label{fig2}
\end{figure}

Based on previous experimental studies~\cite{demidov2009transformation,demidov2011excitation}, variations in the internal effective field near geometrical constrictions—driven by demagnetization effects—can significantly impact spin-wave propagation. To investigate this and ensure proper spin-wave (SW) transmission, crucial for their detection and subsequent characterization with applied current, we first performed PSWS measurements in the absence of current, sweeping the excitation frequency from 4 to 13~GHz under a fixed external magnetic field.

Representative transmitted spectra for a device with \(w_{\mathrm{c}} = 571\)~nm are shown in Fig.~\ref{fig1}(c), displaying the real, imaginary, magnitude, and phase components of the signal. The real and imaginary parts exhibit the oscillatory behavior characteristic of spin-wave phase accumulation between antennas separated by a distance \(D\)~\cite{vlaminck2010spin}. The phase signal varies linearly with frequency within the transmission band and appears noisy outside resonance, consistent with expectations for a dispersive medium. This linear frequency dependence arises because SWs excited at different frequencies propagate at different phase velocities—unlike in nondispersive media such as light in vacuum, where phase remains constant across frequencies. Notably, this phase behavior also encodes information about the group velocity, which can be extracted by analyzing the frequency-dependent phase accumulation, as discussed later.


To understand how geometrical constrictions influence spin-wave propagation, we first extract the dispersion relations for both a reference (unconstricted) strip and a constricted device. As discussed earlier, demagnetization effects in narrow geometries locally modify the internal effective field, potentially suppressing spin-wave transmission at low external fields~\cite{demidov2009transformation,demidov2011excitation}. Figures~\ref{fig2}(a,b) show the experimentally measured dispersions for the reference and $w_c=~$571~nm waveguides, respectively. In the wide device, spin waves are supported down to lower external fields, while in the constricted geometry, a threshold field is required before spin-wave propagation becomes visible. This confirms that the constriction acts as a magnetic barrier for spin waves, and that sufficiently strong external fields are needed to overcome the local demagnetizing effects and enable spin-wave guiding. To further support our interpretation, we performed micromagnetic simulations under comparable conditions (see Supplementary Information), which reproduce the experimental dispersions and the effect of the constriction.


Figure~\ref{fig2}(c) summarizes the minimum magnetic field required for detectable spin-wave transmission ($\mu_0H_{\mathrm{c}}$), extracted from both experiments and simulations as a function of the constriction width $w_{\mathrm{c}}$. Experiment and simulation are in excellent agreement, confirming that demagnetization effects set the threshold field for spin-wave propagation. In narrow constrictions, demagnetization induces a slight canting of the magnetization near the constriction, which, although small relative to the saturation magnetization, leads to partial reflection of spin waves. In contrast, wider constrictions confine this canting to the edges, allowing Damon-Eshbach mode propagation through the central region. As the external field is increased, the magnetization becomes more uniform, reducing demagnetization effects and enabling transmission through narrower constrictions. At low fields, however, suppressed magnetic ordering near the constriction inhibits the propagation of magnetostatic surface waves, effectively creating a transmission stop-band~\cite{vilsmeier2024spatial}.

\begin{figure} 
\centering
\includegraphics[width=\linewidth]{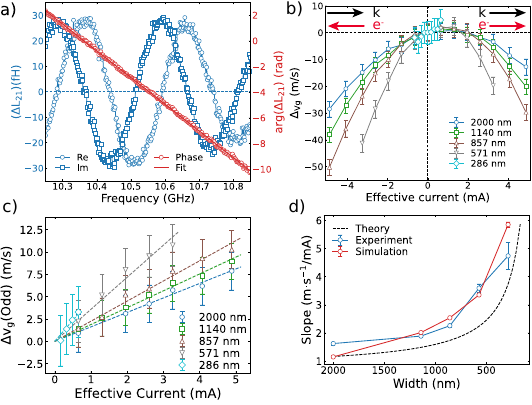}
\caption{\label{fig_3} (a) Real, imaginary, and phase components of the mutual inductance for the sample with $w_{\mathrm{c}} = 286$~nm at $I \approx 0.6$~mA (the noisiest dataset). (b) Absolute change in group velocity, $\Delta v_{\mathrm{g}} = v_{\mathrm{g}}(I) - v_{\mathrm{g}}(0)$, as a function of effective current for various constriction widths. Arrows indicate the relative directions of electron flow ($e^-$) and spin-wave propagation ($k$). (c) Antisymmetric component of the relative group velocity change, with linear fits shown as dashed lines. (d) Slopes of the experimental data fits in (c), from simulations and coefficient of Eq.~\eqref{eq:relativevg} with $M_s=750 \mathrm{~kA}$, $P=0.55$, $g=2.19$, $l=4.5~\mathrm{\mu m}$, $D=6.1~\mathrm{\mu m}$ and $w_0=2000~$nm  highlighting the dominant current-induced contribution to $\Delta v_{\mathrm{g}}$ for narrower constrictions (see Supplementary Material for details on simulated slopes and $P$ value extraction).
}
\end{figure}

As introduced, in a uniform current density distribution, \(v_{\mathrm{g}}\) increases when the spin-polarized electron flow and the wavevector are parallel, due to the uniform STT acting along the propagation path. In a non-uniform current density distribution—such as that created by a constriction—this effect is expected to be amplified, as regions of higher current density locally enhance the STT, leading to stronger modulation of the spin-wave group velocity. To test this hypothesis, we applied a DC current to devices with different constriction widths, focusing on the accumulated phase and the resulting group velocity, \(v_{\mathrm{g}}\). Figure~\ref{fig_3}(a) shows the real and imaginary components of the measured spectra, together with the extracted phase for a representative sample under applied current. The group velocity is obtained by first calculating the group delay time, \(\tau_{\mathrm{g}}(\omega) = -\frac{\mathrm{d}\phi(\omega)}{\mathrm{d}\omega}\), which represents the time taken by the wavepacket envelope to travel from the input to the output antenna, where \(\phi(\omega)\) is the phase as a function of frequency. Knowing the propagation distance (\(D = 6.1~\mu\mathrm{m}\) in our experiments), and following the procedure of Ref.~\cite{WEISS2023170103}, we extract the group velocity as $
v_{\mathrm{g}} = \frac{\partial \omega}{\partial k} = \frac{D}{\tau_{\mathrm{g}}}.$


This method was also used to extract $v_\mathrm{g}$ for the wide strip, with the results shown on Fig.~\ref{fig2}(a). The obtained values are in excellent agreement with previously reported group velocities in Permalloy microstrips~\cite{haidar2012role}. To be able to characterize samples with narrower constriction (e.g. $w_\mathrm{c}=$ 286~nm), we apply an external magnetic field of $\mu_0H = 100$~mT to minimize demagnetizing effects and ensure efficient SW transmission (see Supplementary Material for amplitude data).

Figure~\ref{fig_3}(b) shows the group velocity variation as a function of effective current, assuming 95\% current flow through the ferromagnetic layer due to shunting effects of seeding and capping layers~\cite{gnoatto2025investigating}. For small currents, the response is linear, transitioning to a nonlinear regime at higher currents. We attribute this behavior to two primary contributions: a quadratic component due to Joule heating~\cite{gladii2016spin,gladii:tel-01724624}, and a linear Doppler-like term arising from the spin-drift velocity $v_{\mathrm{d}}$~\cite{SW_doppler_exp}.

To model the interaction between SWs and nonuniform current distributions, we performed micromagnetic calculations in comparable conditions, where Poisson solver for the current distribution was implemented within the Mumax$^3$ simulation platform~\cite{Leliaert_2018}. Details on the implementation can be found in Ref.~\cite{PhysRevApplied.22.054056} and Supplementary Material. All experiments and simulations are performed at $\mu_0H = 100$~mT, where the magnetization is saturated and demagnetizing effects are minimized, as previously discussed.

To interpret the linear regime observed at small currents in Fig.~\ref{fig_3}(b), we developed an analytical model based on the spatially varying spin-wave phase induced by the current gradient. Under the assumption of weak damping and adiabatic propagation, we obtain the following expression for the relative change in group velocity:
\begin{equation}\label{eq:relativevg}
    \Delta v_{\mathrm{g}} = v_\mathrm{d}\left(1 + \frac{l}{2D}\left(\frac{w_0}{w_c} - 1\right)\right)\,,
\end{equation}

\noindent where $l$ is the length of the constriction and $w_0$ the initial width. A detailed derivation, including the effects of finite antenna width, smooth current profiles, and Joule heating, is provided in the Supplementary Material.

To better isolate current-induced effects from quadratic Joule heating and facilitate direct comparison with simulations and theory, Fig.~\ref{fig_3}(c) shows the odd component of the group velocity variation, defined as $\Delta v_{\mathrm{g(odd)}} = \left(\Delta v_{\mathrm{g}}(+I) - \Delta v_{\mathrm{g}}(-I)\right)/2$, for devices with and without constrictions. As expected, narrower constrictions result in larger group velocity modulation for a given current, as the SWs will experience an effective larger current density while they travel between the antennas. In Fig.~\ref{fig_3}(d), we plot the slope of the experimental curves from panel (c), together with the corresponding values obtained from micromagnetic simulations that account for realistic, non-uniform current density distributions (see Supplementary Material). These are compared with the theoretical scaling predicted by Eq.~\eqref{eq:relativevg}. The agreement across experiment, simulation, and theory confirms the geometric dependence of the spin-wave response to spin-polarized current gradients.

In summary, we have shown that inhomogeneous spin-transfer torques, engineered through simple geometrical constrictions, enable deterministic control of the phase and group velocity of propagating spin waves. Our combined experimental, numerical, and analytical analysis demonstrates that current-induced gradients directly modulate the dispersion, providing a scalable route to phase control under fixed applied current. This functionality establishes a foundation for reconfigurable wave-based devices such as magnonic interferometers, while also offering a new experimental pathway toward analogue-gravity phenomena in magnetic systems, where spatially varying spin-drift velocities emulate curved spacetime.

We thank Bert Koopmans for valuable discussions. This work was supported by the Dutch Research Council (NWO), the Research Foundation-Flanders (FWO), and the Special Research Funds of the University of Antwerp (BOF-UA). The computational resources used in this work were provided by the VSC (Flemish Supercomputer Center), funded by Research Foundation-Flanders (FWO) and the Flemish Government -- department EWI.



\clearpage


\bibliography{references}

\end{document}


\preprint{APS/123-QED}

\title{SUPPLEMENTARY: Controlling Spin-Waves by Inhomogeneous Spin-Transfer Torques}
\author{Lorenzo Gnoatto*}
\affiliation{Department of Applied Physics and Science Education, Eindhoven University of Technology, P.O. BOX 5132, 5600 MB Eindhoven, The Netherlands\\}
    \email{l.g.gnoatto@tue.nl}
\author{Jean F. O. da Silva}
\affiliation{COMMIT, Department of Physics, University of Antwerp, Groenenborgerlaan 171, B-2020 Antwerp, Belgium\\}

\author{Artim L. Bassant}
\affiliation{Institute for Theoretical Physics, Utrecht University, Princetonplein 5, 3584 CC Utrecht, The Netherlands\\}

\author{Rai M. Menezes}%
\affiliation{COMMIT, Department of Physics, University of Antwerp, Groenenborgerlaan 171, B-2020 Antwerp, Belgium\\}
\affiliation{Departamento de Física, Universidade Federal de Pernambuco, Cidade Universitária, 50670-901 Recife-PE, Brazil\\}

\author{Rembert A. Duine}
\affiliation{Department of Applied Physics and Science Education, Eindhoven University of Technology, P.O. BOX 5132, 5600 MB Eindhoven, The Netherlands\\}
\affiliation{Institute for Theoretical Physics, Utrecht University, Princetonplein 5, 3584 CC Utrecht, The Netherlands\\}

\author{Milorad V. Milošević}%
\affiliation{COMMIT, Department of Physics, University of Antwerp, Groenenborgerlaan 171, B-2020 Antwerp, Belgium\\}

\author{Reinoud Lavrijsen}
\affiliation{Department of Applied Physics and Science Education, Eindhoven University of Technology, P.O. BOX 5132, 5600 MB Eindhoven, The Netherlands\\}

\date{\today}

\maketitle


\section{\label{sec.EXPERIMENTAL}EXPERIMENTAL}
\subsection{Amplitude measurements}
\begin{figure}[ht]
\centering
\includegraphics[width=1\linewidth]{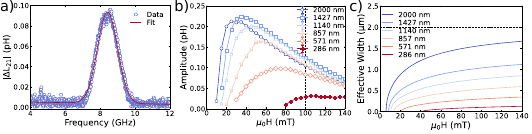}
\caption{a) Transmitted spin-wave spectrum for a device with constriction width $w_{\mathrm{c}} = 571$~nm under an applied field of $\mu_0 H = +60$~mT, along with a Gaussian fit. b) Extracted fit amplitudes for all devices with varying constriction widths. c) $w_{eff}$ calculated for differend widths as a function of the applied magnetic field, black dashed line indicate the nominal 2000 nm width.}
\label{exp.fig.1}
\end{figure}
Figure~S\ref{exp.fig.1}(a) shows a representative transmitted signal fitted with a Gaussian function. This dataset corresponds to the same device shown in Fig.~2(c) of the main text (though at a different external magnetic field), and reflects the spin-wave (SW) amplitude transmitted from antenna 1 to antenna 2. As discussed in the main text, the geometrical constriction modifies the internal effective field, acting as a potential scattering or damping site for SWs due to local variations in demagnetizing fields.

To quantify the overall transmission, we extract the amplitude of the fitted Gaussian at each value of the applied field. The resulting amplitudes, summarized in Fig.~S\ref{exp.fig.1}(b), show a clear trend: the signal amplitude initially increases with field, reaches a maximum, and subsequently decreases. This behavior arises from the interplay between two key mechanisms: the field-dependent effective width \( w_{\mathrm{eff}} \) of the waveguide and the corresponding spin-wave group velocity \( v_{\mathrm{g}} \).

The effective width, \( w_{\mathrm{eff}} \), accounts for demagnetizing effects and determines the onset of SW conduit behavior. Following the approach in Ref.~\cite{bracher2017creation}, we calculate \( w_{\mathrm{eff}} \) and plot it as a function of external magnetic field in Fig.~S\ref{exp.fig.1}(c). At low fields, \( w_{\mathrm{eff}} \) is significantly reduced due to strong demagnetizing effects near the edges, suppressing spin-wave propagation and detection. As the field increases, the magnetization aligns more uniformly along the external field direction, effectively broadening the conduit region. This results in increased excitation and transmission of Damon-Eshbach spin waves, enhancing signal detection.

Once the strip becomes fully magnetized—marked by the maximum in transmitted amplitude—further increasing the external field leads to a gradual decline in signal. This behavior is attributed to the reduction in group velocity \( v_{\mathrm{g}} \), as shown in Fig.~2(a) of the main text. Although the conduit condition is maintained, the lower \( v_{\mathrm{g}} \) at higher fields reduces the efficiency with which spin waves propagate between the antennas, resulting in fewer SWs reaching the detector and, consequently, a weaker measured signal.

Note that the \( w_{\mathrm{eff}} \) model shown in Fig.~S\ref{exp.fig.1}(c) applies to a uniform strip and does not capture spatial variations present in constricted geometries. Nonetheless, the same qualitative trend is observed across all devices. In particular, the maximum signal amplitude decreases systematically with decreasing constriction width \( w_{\mathrm{c}} \), consistent with increased internal field inhomogeneity and enhanced reflection in narrower constrictions. Based on these results, we identify \( \mu_0 H = 100~\mathrm{mT} \) as the minimum field required to ensure reliable SW transmission across all tested constriction widths.

\begin{figure}[ht]
\centering
\includegraphics[width=1\linewidth]{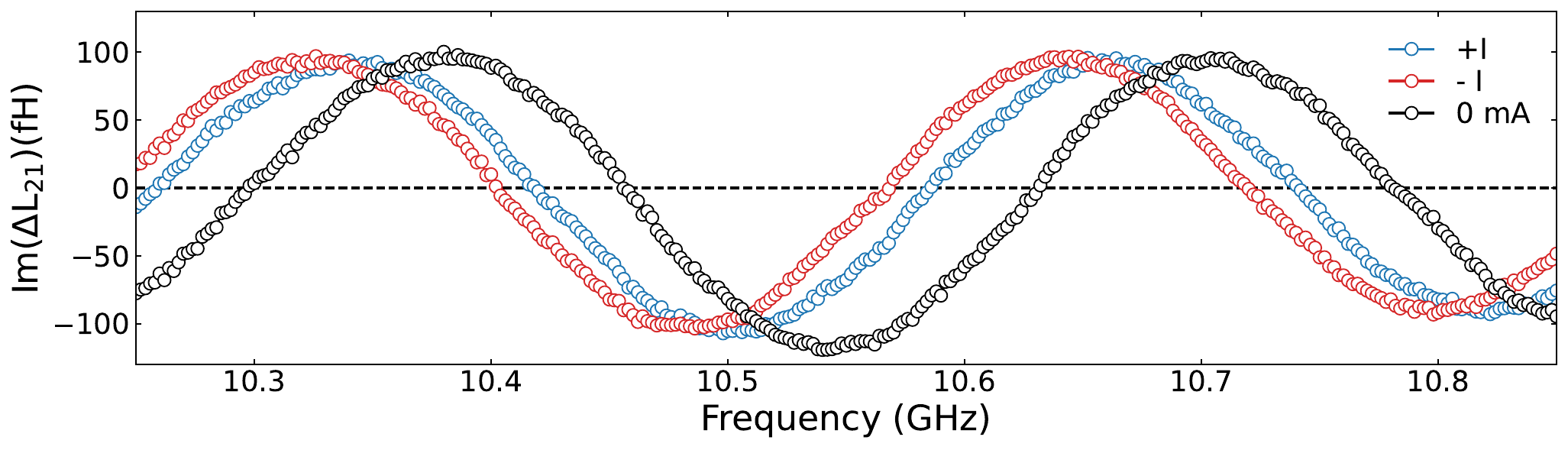}
\caption{Example of a transmitted spectrum for a sample with $w_{\mathrm{c}} = 857\,\mathrm{nm}$, shown at zero current and under applied currents of approximately $\pm5\,\mathrm{mA}$, with an external magnetic field of $100\,\mathrm{mT}$.}
\label{exp.fig.03}
\end{figure}

Figure~\ref{exp.fig.03} shows an example of transmitted spectra and its modification under applied current. When a current is applied, an overall shift toward lower frequencies is observed. 

\begin{figure}[ht]
\centering
\includegraphics[width=1\linewidth]{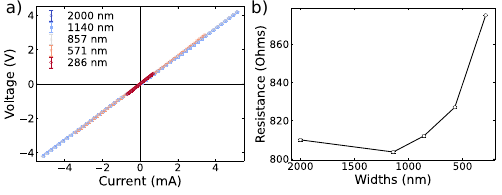}
\caption{2-point measured transport characteristics of all devices: a) IV curves of all devices and (b) resistance as function of constriction width}
\label{exp.fig.02}
\end{figure}




\subsection{Spin polarization extraction using current-induced spin wave Doppler shift}
\label{sub:SWDS}
To determine the degree of spin polarization, we used spin wave Doppler shift measurement on a continuous (unconstricted) strip. The data analysis is the same as our previous report where we determine the Doppler shift as~\cite{gnoatto2024investigating}: 

\begin{align}
\Delta f_\text{dop} &= \frac{\delta f_{\rm 12} - \delta f_{\rm 21}}{4} 
= -\frac{g \mu_\text{B} P}{4 \pi M_{\rm s} |e|} \frac{I_{\text{FM}}}{w\ t}k, \label{eq:doppler_frequency}
\end{align}
where $\delta f_{\rm 12}$ and $\delta f_{\rm 21}$ are the current induced frequency shift for oppositely propagating spin waves, $g\approx2$ is the g-factor, $\mu_{\text{B}}$ the Bohr magneton, $M_{\rm s}$ = 750 kA/m the saturation magnetization, $e$ the electron charge, $k=4.9~\mathrm{\mu m^{-1}}$ the spin-wave wavevector dictated by the antenna, $w$ = 2000 nm is the width of the ferromagnetic strip, $I_{\text{FM}}$ the current through the ferromagnet and $t$ the nominal film thickness, resulting on a value of $P$ = 0.55 $\pm ~0.07$ that we use to plot the theoretical formula in Fig.~3(d) in the main text.


\section{\label{sec.Simulations}Simulations}

\begin{figure}[]
\centering
\includegraphics[width=0.5\linewidth]{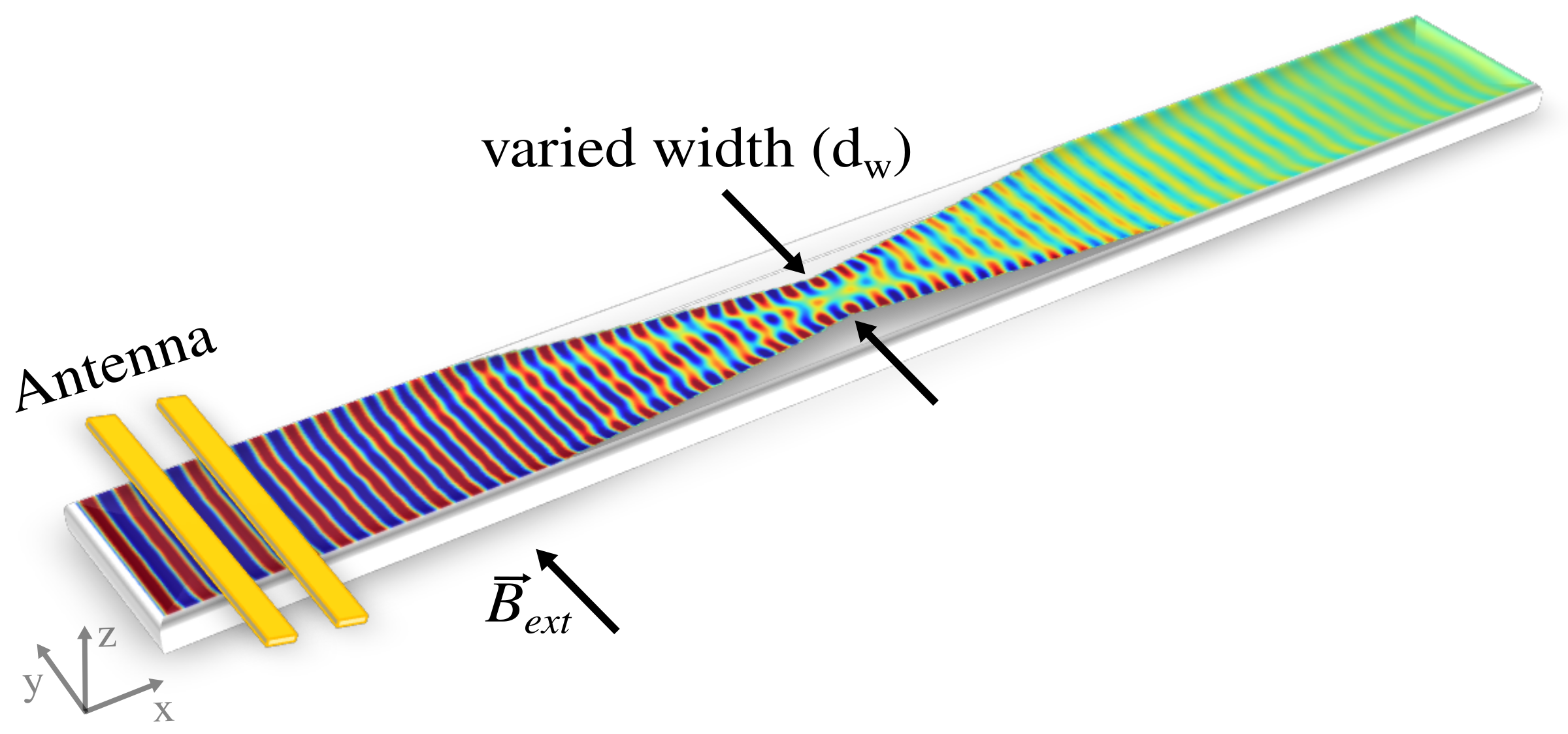}
\caption{Spin-wave propagation in a strip with a constriction in its center. The SW is generated in the antenna in +$z$-direction as a result of an oscillating field $b(t) = b_0\text{sinc}(2\pi ft)$ and propagates in +$x$-direction. $B_{ext}$ is applied in +$y$-direction. Different constriction widths $d_{w}$ are considered to study their effect on the SW propagation.}
\label{sim.fig.1}
\end{figure}

To simulate the considered magnetic system, the Mumax$^3$ package was employed \cite{Vansteenkiste2014,Leliaert_2018}. The magnetization of the ferromagnetic film is represented by the vector field $\vec{M}(\bf r)$ = $M_S\vec{m}(\bf r)$, where $\vec{m}(\bf r)$ is the normalized magnetization direction, as $|\vec{M}| = M_S$ for any position $\textbf{r}$ of the film. The considered energy functional takes into account the contribution of exchange interaction, the Zeeman energy due to applied bias magnetic field and the demagnetizing (dipolar) interaction

\begin{figure}[]
\centering
\includegraphics[width=1\linewidth]{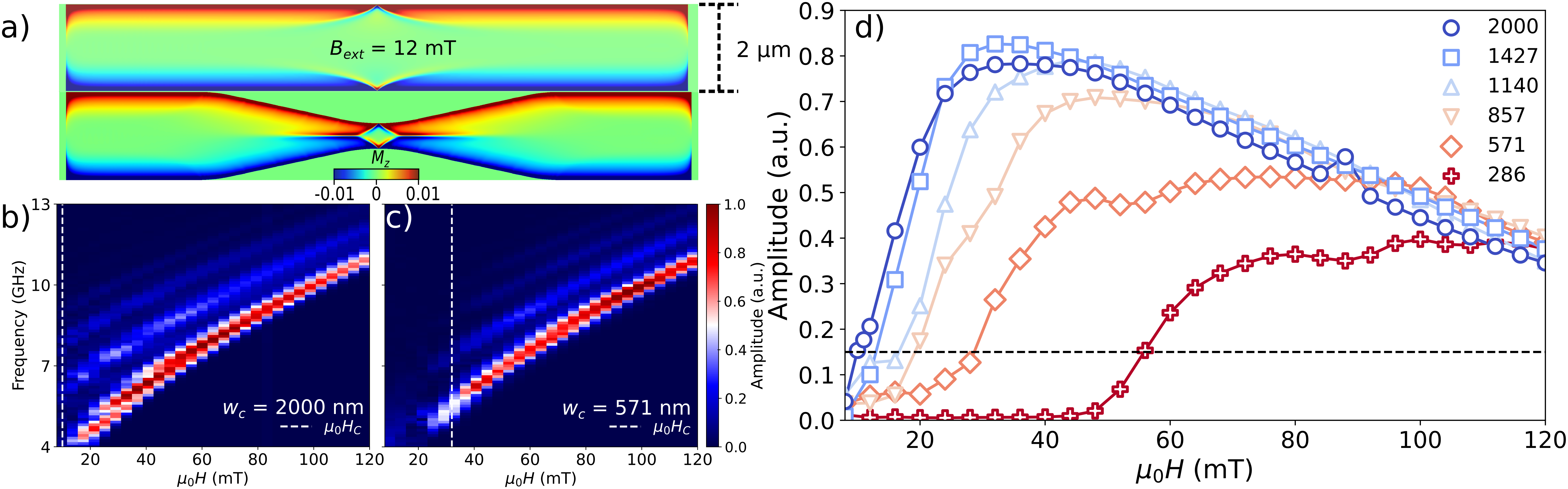}
\caption{\label{sim.fig.4} a) Simulated magnetization distribution for a uniform and a constricted waveguide geometry. Simulated spin-wave intensity at the detection region as a function of applied magnetic field for (b) the reference, unconstricted waveguide (\( w_{\mathrm{c}} = w= 2000~\mathrm{nm} \)) and (c) a constricted waveguide (\( w_{\mathrm{c}} = 571~\mathrm{nm} \)). d) Integrated amplitude for different constrictions, the black dashed line denotes the threshold taken to define the critical field.}

\end{figure}

\begin{equation}\label{eq.E}
    E[\vc{m}] = \int \left[ A(\nabla\vc{m})^2 - M_s \vc{B}_{ext} \cdot \vc{m} - \frac{M_s}{2} \vc{B}_\text{demag}  \cdot \vc{m}\right] d^3\bf{r},
\end{equation}
with the exchange stiffness $A>0$, the external field $\vec{B}_{ext}$, and the demagnetization field $\vec{B}_\text{demag}$, respectively. Following the experimental configuration, we consider the external field in the $+\hat{y}$-direction. We create a localized perturbation in the edge of the sample in the $+\hat{z}$-direction that allows us to generate the spin-wave (SW) solutions (see Fig.~S\ref{sim.fig.1}). Since the perturbation is small, we can neglect the oscillations in $m_y$, thus $\delta \vc{m} = [\delta m_x, 0, \delta m_z]$. The magnetization dynamics is governed by the Landau-Lifshitz-Gilbert (LLG) equation \cite{gilbert2004phenomenological}:

\begin{equation}
    \dot{\vec{m}} = -\gamma \vec{m} \times \vec{H}_{eff} + \alpha \vec{m} \times \dot{\vec{m}} + \vec{\tau}_{STT}, \label{LLG_equation}
\end{equation}
where $\gamma$ is the gyromagnetic ratio, $\vec{H}_{eff}$ is the effective field given by $\vec{H}_{eff} = -\delta E/\delta \vec{M}$, and $\alpha$ is the Gilbert damping factor. The last term on the right-hand side of Eq.~\eqref{LLG_equation} represents the spin transfer torque $\vec{\tau}_{STT}$, which arises when spin-polarized currents are present \cite{RALPH20081190, PhysRevB.70.024417, PhysRevLett.93.127204}.

For the simulations, we consider a $16000\times2000\times 20$~nm$^3$ Permalloy strip, with saturation magnetization $M_{\rm s}=750$~kA/m, gyromagnetic factor $\gamma/2\pi = 29$~GHz/T, $\alpha = 0.007$ and the exchange stiffness constant $A = 13$~pJ/m. The $M_{\rm s}$ value was taken from the experimental measurements, and the other parameters were taken from literature \cite{Ni_Py_parameters,exchange_stiffness_parameters_Py}. The system was discretized into $5\times5\times5$~nm$^3$ cells. 

To generate the SWs in the simulations, we consider an artificial antenna to create a localized perturbation with the field $b(t) = b_0\text{sinc}(2\pi f_{max}t) \hat{z}$ applied along the $z$-direction, with amplitude $b_0 = 0.01B_\text{ext}$ and oscillation frequency ranging from 0 to $f_{max}$ = 15 GHz. This creates spin wave solutions that propagate towards the $+\hat{x}$-direction, as illustrated in Fig.~S\ref{sim.fig.1}, to evaluate the transmitted SW characteristic as a function of the external field $B_{ext}$.

Figure~S\ref{sim.fig.4} shows the micromagnetic simulations for the spin-wave dispersion, performed under conditions similar to those used in the experiments (see Figure 2(a-b) in the main text). In Fig.~S\ref{sim.fig.4}(a) we present the magnetization profile, here, the system is first relaxed to obtain the ground-state magnetization configuration. Because of the geometry and the resulting demagnetizing fields, this relaxation produces a multidomain state, most prominently at the constriction center where the $m_\mathrm{x}$ component exhibits symmetric dips. After applying the transverse field to generate the Damon–Eshbach mode, the domain pattern persists, particularly near the sharp corners of the constriction. This residual nonuniformity does not qualitatively affect the results, and all simulated trends remain fully consistent with the experimental observations. Fig.~S\ref{sim.fig.4}(b) shows the dispersion relation for the full stripe and Fig.~S\ref{sim.fig.4}(c) for the constricted geometry ($w$ = 571 nm). Figure ~S\ref{sim.fig.4}(d) shows the integrated amplitude for the full spectra for each $B_{ext}$, considering all the constrictions, where the black dashed line sets the threshold to define the critical field. These spectra were extracted by recording the transmitted amplitude as a function of the applied external magnetic field. The simulated dispersion shows more features compared to experiments. (i) Multiple dispersion branches are observed which we speculate to be from the excitation of higher-order transverse modes due to the broadband nature of the sinc-function excitation. In particular, both \( n = 1 \) and \( n = 3 \) magnetostatic surface wave modes may be excited, each with distinct dispersion characteristics~\cite{10.1063/1.5099358}. Another speculation might be that they are due to the multidomain nature of the magnetization profile, e.g. domain wall guided spin-wave conduction, albeit a very different dispersion would be expected. (ii) In Fig.S\ref{sim.fig.4}(d) extra steps are observed in the amplitude most prominet in the 571 and 286 nm constrictions, these are due to the dissapearance of the multidomain structure at the constriction being expelled with larger external magnetic field, something we do not observe in the experiments. Further analysis and targeted simulations would be required to unambiguously identify the nature of the modes and effect of the multidomain nature on PSWS. Such an investigation lies beyond the scope of this work. In contrast, these higher-order features are typically not resolved in PSWS measurements, likely due to the limited mode selectivity of the antenna geometry, which primarily excites the lowest-order modes~\cite{lucassen2019optimizing}.

Thereafter, to consider the bias of the charge carriers in the magnetization dynamics \cite{SLONCZEWSKI1996L1, PhysRevB.54.9353}, we incorporate the spin transfer torque term $\vec{\tau}_{STT}$ in the LLG equation following previous steps in the literature \cite{PhysRevLett.93.127204,PhysRevB.69.094421}
\begin{equation}
    (\partial_t + \vec{v}_s \cdot \nabla)\vec{m} = -\gamma\vec{m} \times \vec{H}_{eff} + \alpha \vec{m}\times (\partial_t + \frac{\beta}{\alpha}\vec{v}_s\cdot \nabla)\vec{m} \label{equation_doormenbal},
\end{equation}
the equation above can be linearized assuming small damping $\alpha$ and non-adiabaticity $\beta$ to study macroscopic effects of the STT via \cite{menezes2024toward}:
\begin{equation}
    \omega =  \frac{2A\gamma}{M_S}k^2 + \vec{v}_s \cdot \vec{k} + \gamma B_{ext}. \label{disp_STT}
\end{equation}
This equation expresses the macroscopic interaction due to the addition of a polarized current in the system without considering dipolar interactions. The typical parabolic dispersion relation for SWs in the exchange regime is shifted by a factor \(\Delta\omega \approx \vec{v}_s \cdot \vec{k}\). It means the current density will provide a (Doppler) shift in the dispersion relation calculated by $\frac{\Delta \omega}{k} = \frac{g\mu_BPJ}{2|e|M_S}$ for a rectangular geometry. By considering the micromagnetic simulation, one can take a step further, taking into account the current distribution for each constriction. By solving the Poisson equation to consider a realistic current distribution, therein study the STT effect with more precision, considering that due to non-homogeneity in the current distribution, the results may follow a different trend. 

\begin{figure}[h]
\centering
\includegraphics[width=1\linewidth]{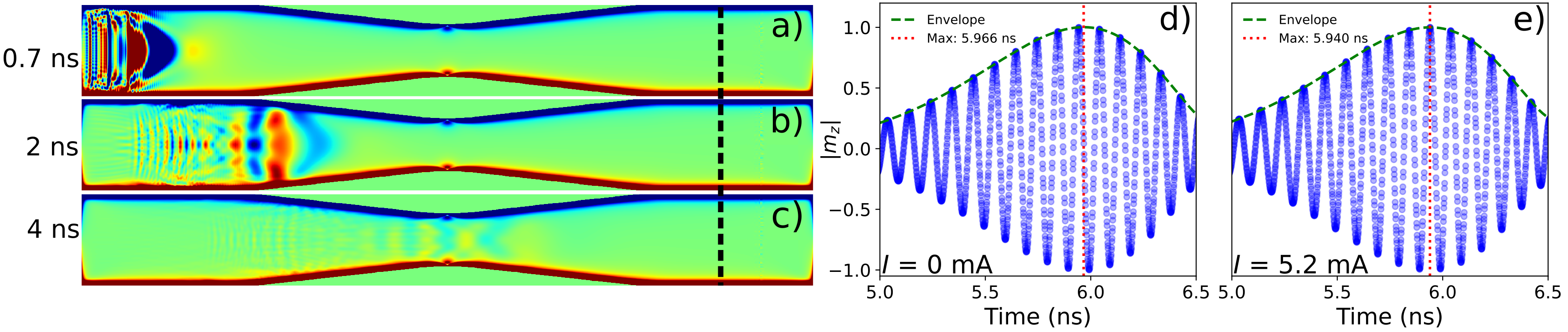}
\caption{Spin-wave propagation in a constricted strip ($w$ = 1140 nm) for (a) 0.7, 2, and 4 ns. The spin wave (SW) reaches the dashed line, where its maximum amplitude is computed for (d) $I$ = 0 mA. The peak amplitude is then analyzed as the current increases to determine the group velocity shift up to (e) $I$ = 5.2 mA.}
\label{sim.fig.3}
\end{figure}

To explore spin-wave dynamics under inhomogeneous current distributions, we extended the simulations to include spatially varying current densities based on realistic geometries. Following the methodology of Ref.~\onlinecite{menezes2024toward}, the current distribution was obtained by solving the Poisson equation and incorporated into the micromagnetic model.

To investigate the dynamics of the SWs under inhomogeneous current distributions, we modeled a 20 nm-thick magnetic thin film, see Fig. S\ref{sim.fig.3}(a-c), and numerically analyzed the distribution of current within the constricted geometry. To isolate the influence of current on SW propagation, a time-dependent perturbation field $b(t) = b_0\text{sin}(2\pi ft) \hat{z}$ (with $f$ = 11 GHz) was applied for 20 ps. Initial simulations were conducted without any applied current to establish a baseline, enabling the extraction of the maximum oscillation amplitudes, see Fig. S\ref{sim.fig.3}(d). We determined the group velocity of the wavefront by fitting an envelope curve within the region of interest in the time domain, delineated by the dashed boundaries in Fig. S\ref{sim.fig.3}(a-c). The propagation distance remained constant (i.e. 13.9 $\mu$m) in all the simulated geometries, thus the calculation of the group velocity is given by $v_g = \frac{\Delta x}{\Delta t}$. We then repeated this analysis, considering different current magnitudes $I$. By tracking the time associated with the maximum amplitude of the envelope, Fig. S\ref{sim.fig.3}e), one can calculate the displacement as the current increases, therefore calculating the group velocity's variation $\Delta v_g = v_g(I) - v_g(I=0)$ for each applied current.

\begin{figure}[]
\centering
\includegraphics[width=0.9\linewidth]{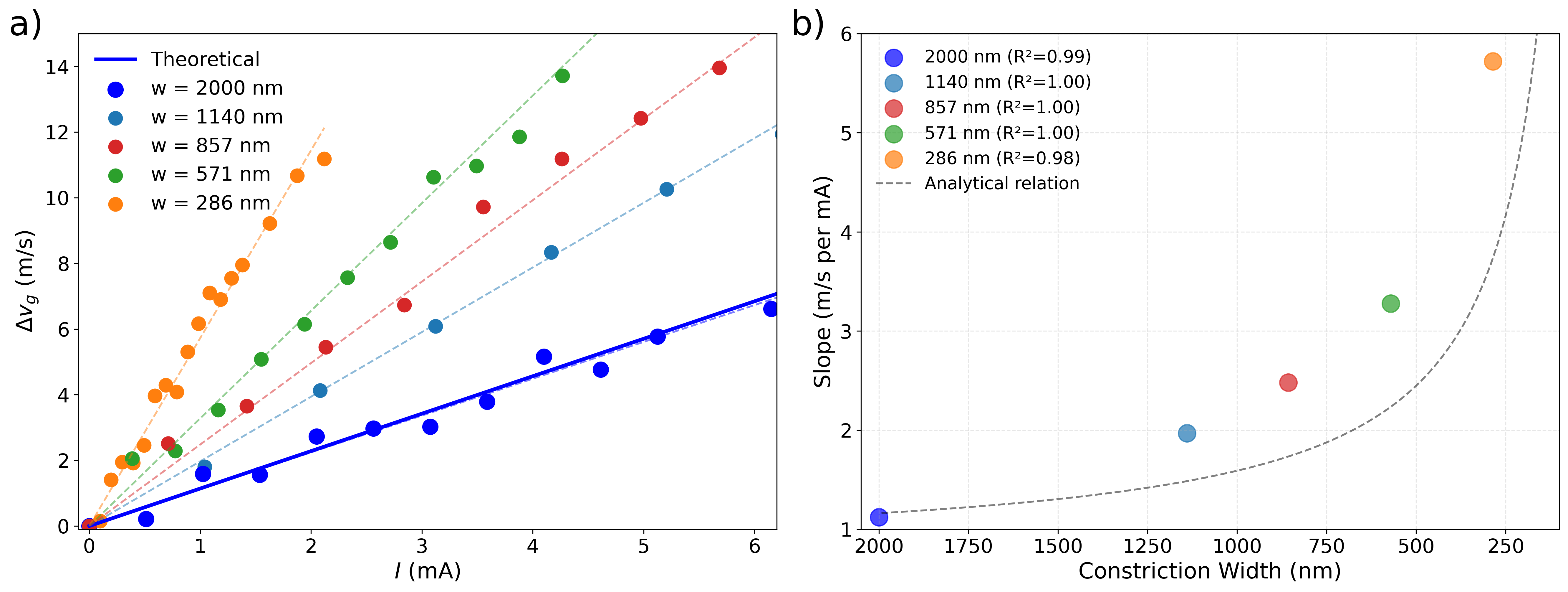}
\caption{a) Simulated variation in spin-wave group velocity \( \Delta v_g \) as a function of current for different constriction widths. b) Extracted slopes of \( \Delta v_g(I) \) compared with theoretical predictions from Equation (1) in the main text.}
\label{sim.fig.5}
\end{figure}

The full result can be seen in Fig. S\ref{sim.fig.5}(a), where one can find the group velocity variation for different constrictions with different current magnitudes. One should notice that the linear behaviour changes as we reduce the width of the constriction.
We took the slopes from Fig.~S\ref{sim.fig.5}(a) to compare with the experimental and analytical results, as shown in the main text.

We also employed simulations to compare these results with the analytical relation developed in the next section. As one can observe in Fig.~S\ref{sim.fig.5}(b), the SW group velocity indicates the same trend as seen in the analytical relation, but the slopes for the simulation results are a bit higher, possibly indicating the dipolar interactions in the sample acting to enhance the SW group velocity.

\section{\label{sec.theory}Analytical theory}

Here we show the derivation of Eq. (1) from the main text. The derivation begins by expressing the inductance as a spatial integral and the susceptibility in terms of spatial coordinates.
Subsequently, we assume small Gilbert damping ($\alpha\approx 10^{-3}\ll1$), employ a simplified antenna model, and consider the spin waves to propagate adiabatically.
This allowed us to extract the phase of the inductance, given by $\phi(\omega)=\int_{x_1}^{x_2}k_\omega(y)dy$ where $x_1$ and $x_2$ are the positions of the antennae and $k_{\omega}(x)$ the spatially dependent wave vector.
Using a model current density given in Eq. \eqref{eq:modelCurrent}, we find an analytical expression for the phase.
The group velocity from the analytical expression of the phase is found using $v_{\mathrm{g}} = D\left(\frac{\mathrm{d}\phi(\omega)}{\mathrm{d}\omega}\right)^{-1}$, which is given in the main text.

Additionally, we discuss the limitations of the analytical result and outline subsequent steps to improve its accuracy.





\subsection{The analytical model of inhomogeneous PSWS}\label{App:groupvelocity}

\begin{figure}[h]
\centering
\includegraphics[width=0.65\linewidth]{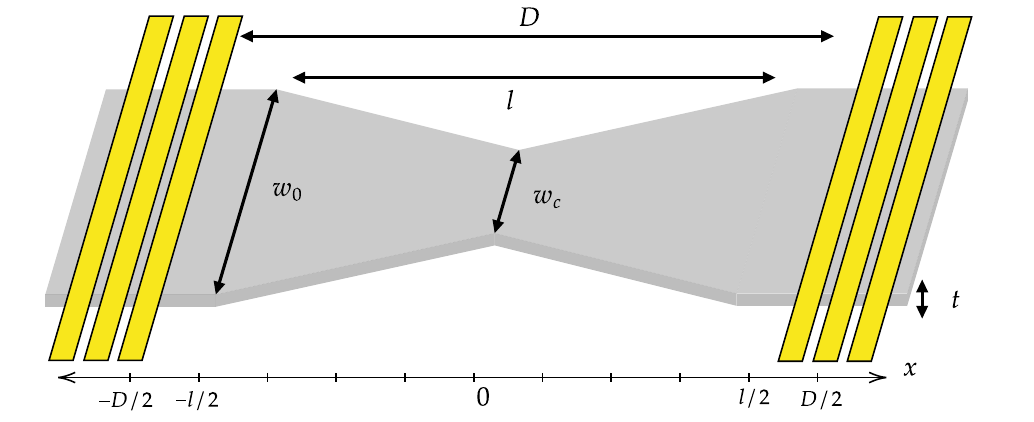}
\caption{A schematic of the setup showing all of the relevant parameters of the geometry.}
\label{fig.1}
\end{figure}

In the following derivation, we will make use of the following parameters of the thin film geometry.
The distance between the antennae is given by $D$, the length of the geometry is given by $l$, the largest (smallest) width of the thin film is given by $w_0$ ($w_c$), and the thickness of the film is given by $t$.
These are schematically given in Fig. S\ref{fig.1}.
In PSWS experiments, the inductance is usually well approximated by the equation \cite{huber_control_2013}:

\begin{equation}\label{eq:Lmom}
    L(\omega)=\frac{\mu_0}{2}\int \rho^2(k)\Tilde \chi(k,\omega)e^{i kD} dk,
\end{equation}

\noindent with $\rho(k)$ the Fourier transform of the magnetic field from the antennae, and $\Tilde \chi(k,\omega)$ is the magnetic susceptibility.
The Fourier transform assumes that the wave guide is translationally symmetric, which is broken by a varying geometry.
Therefore, we do not transform to reciprocal space in deriving Eq. \eqref{eq:Lmom}, and we find

\begin{equation}\label{eq:Lspac}
    L(\omega)=\frac{\mu_0}{2}\iint dx dx' h(x-x_2) h(x'-x_1) \chi(x,x',\omega),
\end{equation}

\noindent where $x_1$ and $x_2$ are the positions of antenna 1 and 2, respectively.
The magnetic field of the antenna in real space is given by $h(x)$, which we assume to be identical for both antennae.
The susceptibility is given by the eigenfunction expansion,

\begin{equation}\label{eq:sus}
    \chi(x,x',\omega)=\int\frac{d\varepsilon}{2\pi}\rho(\varepsilon)\frac{\psi_\varepsilon(x)\psi_\varepsilon^*(x')}{(1+i \alpha)\omega-\varepsilon}.
\end{equation}

\noindent Here, $\psi_\varepsilon(x)$ is an eigenfunction of the linearized Landau-Lifshitz equation with frequency $\varepsilon$, $\rho(\varepsilon)$ is the density of states of spin waves, and $\alpha$ is the damping coefficient.
The eigenfunctions form an orthogonal basis, which allows for this expansion.
Unfortunately, the eigenfunctions are generally difficult to compute for arbitrary geometries, in particular for dipolar spin waves.
We continue by assuming that the spin waves evolve adiabatically as this move through the constriction. 
This entails that the spin waves do not scatter against the boundary and that their wave vector varies smoothly through space.
This results in the eigenfunction $\psi_{\varepsilon}(x)\propto\exp(i\int_{x_1}^x k_\varepsilon(y)dy)$.
Together with the assumption that the Gilbert damping is small, we use the Sokhotski-Plemelj theorem to find the imaginary part of the susceptibility

\begin{align}\label{eq:sus}
    \Im\left(\chi(x,x',\omega)\right)=&\frac{\rho(\omega)}{2}\Im(\psi_\omega(x)\psi_\omega^*(x')), \\
    \propto&\frac{\rho(\omega)}{2}\Im\left(\exp\left({i\int_{x'}^{x}k_\omega(y)dy}\right)\right). \nonumber
\end{align}

\noindent The Kramer-Kronig relation allows us to compute the real part of the susceptibility. 
However, in this case, the imaginary part suffices to find the phase of the non-local inductance.
We model the antennae as a Dirac delta ($h(x-x')=\delta(x-x')$), which equates to an infinitely thin strip.
The generalisation to antennae with finite width does not affect the physical result. 
Varying the antenna’s width changes the distribution of the excited wave vectors, but each wave vector has an equivalent STT term in the dispersion.
The mathematical treatment, however, becomes more involved.
We find that the imaginary part of the inductance is

\begin{equation}
    \Im(L(\omega))\propto\Im\left(\exp\left({i\int_{x_1}^{x_2}k_{\omega}(y)dy}\right)\right).
\end{equation}

\noindent Therefore the phase of the inductance is given by $\phi(\omega)=\int_{x_1}^{x_2}k_\omega(y)dy$.
To compute the phase of the constricted geometry, we need to derive the spatially dependent wave vector $k_\omega(x)$.
For $kt \ll 1$ and $k\approx k_{exp}$ where $k_{exp}$ is the experimentally excited wave vector and $t$ the thickness, we find that 

\begin{equation}
    k_\omega(x)=\frac{c_1+c_2 \omega }{c_3+c_2 c_J j_e(x)}.
\end{equation}

\noindent Herein, we do not take into account the current effects on the magnetisation, such as Joule heating.
Their effects are considered at the end of this section.
The current density is the spatially dependent part of the wave vector and is given by $j_e(x)$.
The coefficients $c_i$ are given by

\begin{subequations}
\begin{widetext}
    \begin{align}
    c_1=&-\left(2 B_0 \gamma  (B_0+M_{eff})+\frac{1}{2} \gamma  t k_{exp} M_{eff} M_s\right), \\
    c_2=& 2 \gamma  \sqrt{B_0 (B_0+M_{eff})-\frac{1}{2} t^2 k_{exp}^2 M_s \left(M_{eff}+\frac{M_s}{2}\right)+\frac{1}{2} t k_{exp} M_{eff} M_s},\\
    c_3=& \gamma  \left(\frac{t M_{eff} M_s}{2}-t^2 k_{exp} M_s \left(M_{eff}+\frac{M_s}{2}\right)\right),\\
    c_J=&\frac{\mu_B P}{ |e| M_s}.
\end{align}
\end{widetext}
\end{subequations}

\noindent We used that $B_0$ is the external magnetic field, $M_{eff}$ is the effective magnetization, $M_s$ is the saturation magnetization, $\gamma$ the gyromagnetic ratio, $\mu_B$ the Bohr magneton, $P$ the polarization, and $e$ the elementary charge.
To solve for the phase, we approximate the current density as

\begin{equation}\label{eq:modelCurrent}
    j_e(x)=\left(1-\frac{w_0}{w_c}\right)\frac{I}{A}\min\left(1,\frac{2|x|}{l}\right)+\frac{w_0}{w_c }\frac{I}{A}.
\end{equation}

\noindent Where $I$ is the total current, $A$ is the surface through which the current passes, $w_0$ is the largest width and $w_c$ is the smallest width in the constriction, and $l$ is the length of the geometry.
The current density is defined such that $x=0$ is the middle of the geometry.
Therefore, the antennae are at $x=\pm D/2$, and the geometry starts/ends at $x=\pm l/2$.
This current density ensures that the total current passing is constant throughout the thin film.
The phase can be separated into two parts, 

\begin{equation}
\phi(\omega)=\underbrace{\int_{-D/2}^{-l/2}k_\omega(y)dy+\int_{l/2}^{D/2}k_\omega(y)dy}_{\phi_0}+\underbrace{\int_{-l/2}^{l/2}k_\omega(y)dy}_{\phi_1}.
\end{equation}

\noindent The total phase is thus given by $\phi=\phi_0+\phi_1$, where $\phi_0$ is the phase accumulated outside the canting geometry and $\phi_1$ is the phase accumulated inside the geometry.
The two parts of the phase can be computed,

\begin{subequations}
\begin{align}
    \phi_0=& (D-l)\frac{c_1+c_2 \omega }{c_3+c_2 c_J j_{min}}, \label{eq:phi0}\\
    \phi_1=&2\int_{j_{min}}^{j_{max}}\left|\frac{dj_e}{dx}\right|^{-1} \frac{c_1+c_2 \omega }{c_3+c_2 c_J j_e}dj_e,  \nonumber\\
    =&\left(\left(\frac{w_0}{w_c}-1\right)\frac{j_{min}}{l}\right)^{-1}\frac{c_1+c_2 \omega }{c_2 c_J}\ln \left(\frac{c_3+c_2 c_J j_{max}}{c_3+c_2 c_J j_{min}}\right). \label{eq:phi1}
\end{align}
\end{subequations}

\noindent The integration is done from the lowest current density $j_{min}=I/A$ to the largest current density $j_{max}=w_0I/w_c A$.
In the experiment, we find that the spin-drift velocity $v_d=c_J I/A$ is smaller than the group velocity of the experimentally excited wave vector in the Damon-Eshbach (DE) geometry, $v_{DE}=c_3/c_2$.
Therefore, we linearised the phase using $v_d\ll v_{DESW}$ or equally $j_{min}\ll c_3/c_2c_J$.

\begin{subequations}
\begin{align}
    \phi_0\approx&(D-l)\frac{c_1+c_2 \omega }{c_3 }\left(1-\frac{c_2 c_J}{c_3 }j_{min}+\frac{c_2^2 c_J^2}{c_3^2}j_{min}^2\right), \\
    \phi_1\approx&l\frac{c_1+c_2 \omega }{c_3 }\left(1-\frac{c_2 c_J}{2 c_3} \left(1+\frac{w_0}{ w_c}\right)j_{min}+\frac{c_2^2 c_J^2}{3 c_3^2} \left(1+\frac{w_0}{ w_c}+\left(\frac{w_0}{ w_c}\right)^2\right)j_{min}^2\right).
\end{align}
\end{subequations}

\noindent The sum of these expressions gives us the total phase.
From Eqs. \eqref{eq:phi0}-\eqref{eq:phi1}, we compute the group velocity.
For small current densities, we find

\begin{align}
    v_g=& D\left(\frac{d\phi}{d\omega}\right)^{-1}, \\
    \approx&\frac{ c_3 }{c_2 }\left(1+\frac{c_2 c_J }{ c_3}\left(1+\frac{l}{2D}\left(\frac{w_0}{ w_c}-1\right)\right)\frac{I}{A}-\frac{c_2^2 c_J^2 }{ c_3^2}\frac{l}{D}\frac{4D-3l}{12D}\left(\frac{w_0}{ w_c}-1\right)^2\left(\frac{I}{A}\right)^2\right).\nonumber
\end{align}

\noindent Here, we again used that $j_{min}\ll c_3/c_2c_J$.
The first term is the group velocity at the experimentally excited wave vector, $v_{DE}=c_3/c_2$.
Thus, the change of the group velocity induced by the current is given by

\begin{equation}\label{eq:vg}
    \Delta v_g\approx \underbrace{c_J\left(1+\frac{l}{2D}\left(\frac{w_0}{ w_c}-1\right)\right)\frac{I}{A}}_{\text{STT}}-\frac{l}{D}\frac{4D-3l}{12D}\left(\frac{w_0}{ w_c}-1\right)^2\frac{c_2  }{ c_3}\left(c_J\frac{I}{A}\right)^2.
\end{equation}

\noindent The first term is a change in group velocity solely due to the inhomogeneous STT.
The second term is proportional to $(v_d/v_{DE})v_d$ and therefore also incorporates the largest order correction due to the shape of the dispersion relation.
In the main text, we only consider the first term since the higher-order terms are small and masked by the Joule heating effect.

Eq. \eqref{eq:vg} is a result of several simplifications, such as vanishing Gilbert damping and infinitely thin antennae.
This causes the analytical result to slightly deviate from experimental data.
In particular, both a large Gilbert damping and an antenna with a finite width cause a larger range of spin waves to contribute to the group velocity.
Taking this into account makes the analytical result more difficult, but does not alter the result, since the first term of Eq. \eqref{eq:vg} is independent of the excited wave vectors.

Another simplification is the model for the current density.
In Eq. \eqref{eq:modelCurrent}, we used that at any point in the permalloy, the total current stays constant.
This causes the current density to vary linearly with position, and in the middle it is not smooth. In reality, the current density varies smoothly through the permalloy, which changes the accumulated phase of the spin waves in the constriction.
This effect gives a small correction to the result, which could slightly increase the change in the group velocity.

Lastly, this analytical result is only true for small currents, since it does not take into account effects caused by Joule heating.
The magnetisation of the permalloy decreases approximately quadratically with current due to heating.
Since the heating changes with position, the magnetisation can change with position.
Taking this into account in the numerical evaluation of the change group velocity showed us a similar trend as Fig. 3(d) of the main text.
Unfortunately, taking into account the exact decrease in magnetisation at every position due to current is complex, since the different devices have different heat dissipation.
A possible solution could be simulating or experimentally measuring the magnetisation change due to current.
This, in turn, can be used to numerically evaluate the phase and, therefore, also the group velocity.

\clearpage

\nocite{*}

\bibliography{references_sup}